\documentstyle[12pt,amstex,amssymb] {article}
\newcommand{\beq}{\begin{equation}}

\newcommand{\eeq}{\end{equation}}
\newcommand{\beqa}{\begin{eqnarray}}
\newcommand{\eeqa}{\end{eqnarray}}
\begin {document}
\parindent=15pt
\begin{flushright}
{\bf US-FT/6-97}
\end{flushright}
\vskip .8 truecm
\begin{center}
{\bf MOMENT ANALYSIS, MULTIPLICITY DISTRIBUTIONS AND CORRELATIONS IN HIGH 
ENERGY PROCESSES: NUCLEUS-NUCLEUS COLLISIONS}\\
\vskip 1.5 truecm
{\bf J. Dias de Deus$^*$, C. Pajares and C. A. Salgado.}\\
\vskip 0.9 truecm
{\it Departamento de Fisica de Part\'{\i}culas, Universidade de Santiago de
Compostela, \\
15706--Santiago de Compostela, Spain}
\end{center}
\vskip 2. truecm

\begin{abstract}
\hspace {30pt}

Cumulant oscillations, or $H_q$ moment oscillations, appear if the KNO 
multiparticle distribution decreases at large $z$, $z\equiv n/ <n>$, faster 
than the exponential, $exp(-D z^\mu)$, with $\mu > 1$. In nucleus-nucleus
interactions this behaviour is related to the limitation in the average number
of elementary central collisions (or average number of strings centrally 
produced), due to the finite number of nucleons involved. Colour deconfinement,
via percolating string fusion, will drastically decrease the fraction of 
centrally produced strings and increase the cut-off parameter $\mu$: Moment 
oscillations will be displaced to smaller $q$ and the width of the KNO 
distribution and forward-backward particle correlations will become 
smaller.
\end{abstract}
\vskip 1. truecm

*) Iberdrola visiting professor. On leave of absence from 
Instituto Superior T\'ecnico, 1096 LISBOA codex, Portugal.


\vskip 0.5cm
February 1997 \\

{\bf US-FT/6-97}

\pagebreak

In the framework of perturbative QCD it was theoretically predicted, sometime
ago, that the factorial cumulants $K_q$ (or, equivalently, the moments $H_q
\equiv K_q/F_q$, $F_q$ being the factorial moments) of the multiparticle 
distribution should present oscillations in sign as a function of $q$ \cite{1}.
The predictions turned out to be confirmed by experiment in $e^+e^-$ 
annihilations \cite{2}, in $p\bar p$ interactions  \cite{3} and in 
hadron-nucleus and nucleus-nucleus collisions \cite{4}. It is now clear that 
such oscillations are present in all known high energy processes.

Recently, in \cite{5}, it was shown that a necessary condition for a particle
distribution $P(n,<n>)$, $<n>$ being the average multiplicity, with positive 
two particle correlation, $K_2 > 0$, to have oscillations in $K_q$ is to 
exist, asymptotically, a KNO distribution $\Psi(z)$, \cite{6}, of the form:

\begin{equation}
 <n>P(n,<n>) @>> {\begin{Sb}
<n>\to \infty \\
z\equiv n/<n>=const.
\end{Sb}} > \Psi(z) @>> {z\gg1} > exp[-Dz^\mu]\ , 
\label{1}
\end{equation}

\noindent
$z\equiv n/<n>$ being the scaling variable, $D$ and $\mu$ positive parameters,
with 
\begin{equation}
\mu > 1.
\label{2}
\end{equation}

\noindent
Most of the existing popular parametrizations (of the Negative Binomial 
Distibution family) do not satisfy (\ref{1}) and (\ref{2}). They behave 
exponentially at large $z$, $\mu=1$, and, not surprisingly, they do 
\underline{not} show oscillations in $q$: $K_q>0$ for all values of $q$.

The importance of the large $z$ behaviour of the distributions to generate the
oscillations can be easily seen from the expression relating $K_q$ to $F_q$
($F_q\equiv <n(n-1)....(n-q+1)>/<n>^q)$) and $K_p$, $p<q$:

\begin{equation}
K_q=F_q-\sum_{r_1,r_2,...} P^q_{r_1,r_2,...} K_{r_1}K_{r_2}...,
\label{3}
\end{equation}
with $r_1+r_2+...=q$, $r_i\geq 1$, $r_{i+1}\geq r_i$ and $r_1<q$. The 
$P^q_{r_1,r_2,...}$ are the positive combinatorial factors associated to the 
partition of an integer $q$ into integers $r_1$, $r_2$, $...$ . The meaning of 
(\ref{3}) is straightforward. To obtain the $q$ particle cumulant $K_q$ one has 
to substract from the $q$ particle factorial moment $F_q$ (integrated $q$ 
particle inclusive density) all the $q-1$, $q-2$,$...$ particle cumulants 
(integrated inclusive particle correlations) in all clustering combinations.
We have, in particular (note that we are using normalized moments, $K_1=F_1=1$),
\beqa
&&K_2=F_2-1 \notag \\ 
&&K_3=F_3-3 K_2-1 \\
&&K_4=F_4-4 K_3-3 K_2^2-6 K_2-1 \notag \\
&&K_5=F_5-5 K_4-10 K_3K_2-10 K_3-15 K_2^2-10 K_2-1. \notag
\label{4}
\eeqa

If the distribution if of exponential type the $F_q$ grow very fast with $q$,
essentially as $q!$, and all the terms in (\ref{3}) and (\ref{4}) 
remain positive
($K_2$ is assumed positive). If at large $z$ the distribution decreases faster 
than the exponential, $\mu > 1$, the $F_q$ grow slower and at some value of
$q$, $K_q$ start becoming negative. But this negative $K_q$ enter in the future
steps, $q+1$, $q+2$,$\dots$ giving positive contibutions to the right hand side
of (\ref{3}) and (\ref{4}). At some stage the $K_q$ become again positive, etc.: 
the oscillations start.

It can be argued that the required supression at large $z$ is not really 
physical, in the sense that at finite energy and limited experimental 
acceptance, distributions are always cut at large $z$ \cite{7}. In $e^+e^-$
annihilations there are theoretical reasons to belive that the oscillations 
are physical \cite{1}. We shall argue here that in nucleus-nucleus collisions,
at least, the oscillations have also a physical origin.

Succesful models that attempt to explain hadron-hadron, h-h, hadron-nucleus, 
h-A, and nucleus-nucleus, A-B, collisions, making use, or not, of basic 
information from $e^+e^-$ annihilations, are multiple scattering models. We
shall take as reference the Dual Parton Model (DPM), \cite{8}, but most of our
results do not depend on detailed features of a particular model.

In any multiple scattering model the distribution $P(n)$ of produced particles
is the result of the superposition of the contributions from elementary 
inelastic collisions. At each elementary collision particles (via string 
formation, for instance) are emitted with a given distribution. There is a 
certain probability of $\nu$ elementary collisions to occur. In general, we
can write
\begin{equation}
P(n)=\sum_{\nu=1}\sum_{n_1,\dots,n_\nu} \varphi(\nu)p(n_1)p(n_2)\dots
p(n_\nu)
\label{5}
\end{equation}

\noindent
with $n_1+n_2+\dots+n_\nu=n$, $p(n_i)$ being the particle distribution from
the i-th elementary collision, $\varphi(\nu)$ the probability distribution for
$\nu$ elementary collisions. The parameter $\nu$ represents, as well, the 
number of intermediate produced objects: pairs of strings, in DPM. In (\ref{5}) 
we have assumed that the formed strings emit independently and that 
fluctuactions in the size of the strings are negligeable, or can be reabsorbed
in the distribution $\varphi(\nu)$.

Let us now introduce the generating function $G(z)$,
\begin{equation}
G(z)=\sum_{n=0}(1+z)^nP(n)
\label{6}
\end{equation}

with $G(0)=1$, such that

\beqa
F_q=\left.{{1}\over{<n>^q}} {{d^qG(z)}\over{dz^q}}\right|_{z=0} \quad,
\label{7}
\eeqa

\noindent
and
\beqa
K_q=\left.{{1}\over{<n>^q}} {{d^qlogG(z)}\over{dz^q}} \right|_{z=0} \ \ .\quad
\label{8}
\eeqa

\noindent
By combinig (\ref{5}) and (\ref{6}) we obtain
\begin{equation}
G(z)=\sum_{\nu=1}\varphi(\nu)g(z)^\nu,
\label{9}
\end{equation}
where $g(z)$ is the generating functioin for the elementary process. Knowing
the elementary generating function $g(z)$ (from $e^+e^-$, string model, Poisson
approximation, etc) and the elementary collision distribution $\varphi(\nu)$
(from multiple scattering combinatories, impact parameter integrations, 
etc) the 
full generating function $G(z)$ can be constructed and the moments computed,
(\ref{9}), (\ref{7}) and (\ref{8}). A summary of results is contained in 
Table 1.

For the average multiplicity $<n>$ and the normalized KNO dispersion $D/<n>$,
where $D^2=<n^2>-<n>^2$ we have,
\begin{equation}
<n>=<\nu>\bar n
\label{10}
\end{equation}

\noindent
and
\begin{equation}
{{D^2}\over{<n>^2}}={{<\nu^2>-<\nu>^2}\over{<\nu>^2}}+{{1}\over{<\nu>}}
{{d^2}\over{\bar n^2}}\ \ ,
\label{11}
\end{equation}

\noindent
where $\bar n$ and $d$ are the average multiplicity and the dispersion of
the elementary process, respectively.

Concerning relation (\ref{11}) an important observation can be made. If 
fuctuactions in the effective number of strings were negligeable, i.e., 
$(<\nu^2>-<\nu>^2)/<\nu>^2\simeq 0$, the second term in the right hand side 
of (\ref{11}) should dominate. The KNO dispersion, as $<\nu> >1$, should then
be smaller than the normalized dispersion of the elementary process (say,
$e^+e^-$ or Poisson distribution). This is against experiment: $D^2/<n>^2$
increases with the complexity of the systems involved, from 0.09 in $e^+e^-$
annihilations \cite{9}, to 0.25 - 0.30 in $p\bar p$ collisions \cite{10},
to larger values in h-A processes, and to 0.8 - 1 in A-B collisions 
\cite{11,12}. The conclusion is that in (\ref{11}) the first term in the right 
hand side, counting the fluctuations in the effective number of strings, is
dominant. This is particularly true in nucleus-nucleus collisions where, for
large A and B, $<\nu>\sim 10^2-10^3$, thus making the second term in the right 
hand side of (\ref{11}) completely negligeable. What we have shown for $D/<n>$
applies to all the cumulants $K_q$ (see Table 1). At least in nucleus-nucleus
collisions the KNO particle ditribution function $<n>P(n,<n>)$ must be very 
close to the KNO string distribution function $<\nu>\varphi(\nu,<\nu>)$, with 
scaling variable $z=n/<n>\simeq \nu/<\nu>$.

It should by now be clear that oscillations in the factorial cumulants $K_q$
(or in the $H_q$ moments), at least in nucleus-nucleus collisions, have nothing
to do with the elementary interactions and cannot be related in a simple way
to some perturbative QCD calculations (which may apply to the elementary 
process). We shall try next to explain the origin of the oscillations in A-B
collisions.

In nucleus-nucleus A-B collisions, if instead of measuring the unconstrained
multiplicity distributions $P(n)$ one measures the distribution at a fixed
impact parameter $b$ (in particular in very central collisions, at $b=0$) the
obtained distribution is totaly different. While in the unconstrained situation
the KNO distribution is wide ($D/<n>\simeq 1$) and roughly independent of
the nuclei, the central KNO distribution is very narrow ($D/<n>\ll 1$) and
strongly dependent on the nuclei envolved, specially on the lightest nucleus
\cite{11}. These differences can be easily understood from (\ref{11}).

If the impact parameter is fixed the fluctuation in the number of strings is 
drastically reduced, $(<\nu^2>-<\nu>^2)/<\nu>^2 \to 0$, and the second term in 
the right hand side of (\ref{11}) dominates:

\beqa
\left.{{D^2}\over{<n>^2}} \right |_{central} \quad = 
{{1}\over{<\nu>_c}}{{d^2}\over{\bar n^2}}\ \ ,
\label{12}
\eeqa

\noindent
where $<\nu>_c$ is the average number of  central elementary collisions. One 
sees that the KNO distribution has to be very narrow ($d/\bar n$ is almost
Poisson-like and $<\nu>_c$ may be very large), and the width depends on 
$<\nu>_c$. The value of $<\nu>_c$ increases with the atomic weight of the 
nuclei, being limited by the atomic weight of lightest nucleus. Independently
of A and B one roughly has \cite{11}

\begin{equation}
<\nu>/<\nu>_c \simeq 1/4
\label{13}
\end{equation}

A good way of parameterising multiplicity distributions, from $e^+e^-$
annihilations to A-B collisions, is by using the generalized gamma function
for the asymptotic KNO function,
\begin{equation}
\Psi(z)={\mu\over\Gamma(\kappa)}\Big[{\Gamma(\kappa+1/\mu)\over\Gamma
(\kappa)}\Big]^{\kappa\mu}z^{\kappa\mu-1}exp\Big(-\Big({\Gamma(\kappa
+1/\mu)\over\Gamma(\kappa)}z\Big)^\mu\Big),
\label{14}
\end{equation}

\noindent
with

\begin{equation}
<z^q>={\Gamma(\kappa+q/\mu)\over\Gamma(\kappa+1/\mu)^q}\Gamma(\kappa)^{q-1},
\label{15}
\end{equation}

\noindent
and, in particular, $<z^0>=<z^1>=1$. The parameters $\kappa$ and $\mu$ have
to be fixed.

We shall now impose the constraint that the large $n$ behaviour of the central
collision particle distribution must be the same as the behaviour of the 
unconstrained distribution:

\begin{equation}
P(n,<n>) @>>{n\to\infty}> P(n,<n>)_{central}
\label{16}
\end{equation}

If the central distribution is also parametrized as (\ref{14}), with parameters
$\mu_c$ and $\kappa_c$, it is clear, from (\ref{16}), that $\mu_c=\mu$.
However, $\kappa_c$ is very different from $\kappa$, as the central distribution
is very narrow. This means, see (\ref{15}) for $q=2$, $<z^2>\simeq 1$ or 
$\kappa_c \gg 1$. We can finally impose (\ref{16}) in  a stronger way, see
(\ref{14}):

\begin{equation}
{\Gamma(\kappa+1/\mu)\over\Gamma(\kappa)}={<\nu>\over<\nu>_c}\ \ .
\label{17}
\end{equation}

\noindent
In order to have estimates for $\kappa$ and $\mu$ we introduce the additional 
relations, see (\ref{15}),

\begin{equation}
{\Gamma(\kappa+2/\mu)\over\Gamma(\kappa+1/\mu)}{\Gamma(\kappa)\over
\Gamma(\kappa+1/\mu)}={D^2\over<n>^2}+1 \ \ ,
\label{18}
\end{equation}

\noindent
and make use of experimental information on $D/<n>$. Reasonable values for
$\kappa$ and $\mu$ in agreement with \cite{17} and \cite{18}, $<\nu>/<\nu>_c
\simeq 1/4$ and $D/<n>\simeq 0.9$, are:

$$
\kappa \simeq 0.1
$$
and
\begin{equation}
\mu\simeq 5
\label{19}
\end{equation}

A few remarks can now be made:

\ \ \ 1) As, from (\ref{19}), $\kappa\mu<1$, the KNO function does not turn to 
zero as $z\to 0$, in agreement with experimental data on multiplicity and 
transverse energy distributions, \cite{12,13};

\ \ \ 2) The parameter $\mu$ is very large, which means that the particle 
distribution is drasticaly cut at large $n$, in agreement with multiple 
scattering models \cite{11,12,13,14} and data;

\ \ \ 3) As $\mu$ is large, oscillations occur in the cumulants $K_q$.

In Fig.1 we show a plot of the KNO distribution which agrees with experimental
data. In Fig.2 we present $H_q$ as a function of $q$. In both cases $\kappa$ 
and $\mu$ were fixed at the values (\ref{19}).

Let us next suppose that the road to the formation of the quark-gluon plasma
in nucleus-nucleus collisions is by string fusion \cite{15}, mainly occuring
in central collisions, where the density of strings is higher. The net 
result is that the large $n$ tail of the multiplicity distribution is further
cut at large $n$, or, in other words, $\mu$ is further increased. This can 
directly be seen in (\ref{17}): an increase of $<\nu>/<\nu>_c$, due to string
fusion, is translated in an increase in $\mu$.

In Fig.1 and 2 we also show the KNO multiplicity distribution and moments $H_q$
as function of $q$ when the ratio $<\nu>/<\nu>_c$ is increased from roughly
$1/4$ to $1/2$. The oscillations tend to start earlier.

The same kind of effect occurs if one is triggering on a heavy particle
like the $J/\Psi$. The n-tail of the multiplicity distribution is further cut
and therefore the oscillations tend to start earlier, as it is seen in 
Fig.2 where it is ploted $H_q$ for the multiplicity distribution in S-U
collisions at $\sqrt{s}=19.4 \ GeV$ when a $J/\Psi$ is triggered. Also,
It is shown the result when Drell-Yan pairs are triggered instead of
$J/\Psi$.

We look now to the forward-backward correlations. If one fixes the number of
particles forward, and studies the distribution in the backward hemisphere,
one approximately has

\begin{equation}
<n_B>=a+b n_F\ \ ,
\label{20}
\end{equation}
with
\begin{equation}
b\equiv D^2_{FB}/D^2_F\ \ ,
\label{21}
\end{equation}

\begin{equation}
D^2_{FB}\equiv <n_B n_F>-<n_B><n_F>\ \ ,
\label{22}
\end{equation}

\noindent
and $D^2_F$ being the variance in the forward hemisphere. Within the spirit of
our approximation of keeping only fluctuations in the number of elementary 
collisions, we assume that particles are produced symmetrically in center
of mass rapidity, see \cite{18}, to obtain

\begin{equation}
D^2_{FB}={<\nu^2>-<\nu>^2\over<\nu>^2}<\nu_B><\nu_F>
\label{23}
\end{equation}

\noindent
and, see (\ref{11}),

\begin{equation}
D^2_F=\Big[{<\nu^2>-<\nu>^2\over <\nu>^2}+{2\over<\nu>}{d^2\over\bar n^2}
\Big]<n_F><n_F>\ \ ,
\label{24}
\end{equation}

\noindent
where the factor $2$ in the second term in the right hand side of (\ref{24}) 
corresponds to particles being independently emitted in rapidity in each 
elementary collision.

It is clear that in an unconstrained nucleus-nucleus, A-A, collision, from
(\ref{20}), (\ref{21}) and (\ref{22}), the forward-backward correlation parameter
$b$ is essentially $1$. If we trigger in a central collision, $<\nu^2>-<\nu>^2
/<\nu>^2\to 0$, as seen before, and the parameter $b$ drops to zero. For
instance, in the case of percolating fusion \cite{14}, as $<\nu>/<\nu>_c
\simeq 1/2$, if one triggers on forward events with

\begin{equation}
n_F\gtrsim 2<n_F> 
\label{25}
\end{equation}

\noindent
one should already obtain $b\simeq 0$. This agrees with the results of
\cite{20}, but as we have here strong fusion the net effect is more spectacular.

{\bf Acknowledgements}

We thank CICYT of Spain for financial support and one of us (C.A.S.) to 
Xunta de Galicia for a fellowship. We also thank prof. M. A. Braun who
participated in the initial stage of this research.

\pagebreak

\newpage

\begin{center}
{\bf Figure captions}\\
\end{center}

\vskip 1cm
\noindent
{\bf Fig.1}. KNO negative particle distributions in heavy nucleus-nucleus 
collisions.
Continuous curve: present energy situation, $\kappa=0.1$, $\mu=5$ in eq.
(\ref{14}). Dashed curve: the same distribution for percolating string fusion,
with $\kappa=0.1$, $\mu=10$,
corresponding to about 50\% reduction in the density of central strings.

\vskip 1cm
\noindent
{\bf Fig.2}. The moments $H_q\equiv K_q/F_q$ for $q=2, 3, 4, 5, 6, 7$ in 
the case of
nucleus-nucleus collisions without fusion ($\mu=5$, black circles), and with
fusion ($\mu=10$, open circles). Also represented with crosses (squares) the
$H_q$ deduced from the multiplicity distributions of S-U collisions at
$\sqrt{s}=19.4$ GeV when a $J/\Psi$ (a Drell-Yan pair) is triggered.

\pagebreak

\vskip 1cm
\begin{table}[h]
\begin{center}
\begin{Large}
\begin{tabular}{|c||c|}\hline
q & $K_q$ \\
\hline
& \\
2 & $
 \Big[{<\nu^2>-<\nu>^2\over <\nu>^2}\Big]+{\kappa_2\over<\nu>}$\\
 & \\
\hline
& \\
3 & $\Big[{<(\nu-<\nu>)^3>\over <\nu>^3}\Big]+3{<\nu^2>-<\nu>^2\over <\nu>^2}
{\kappa_2\over<\nu>}+{\kappa_3\over<\nu>^2}$ \\
& \\
\hline
& \\
 & $ \Big[{<(\nu-<\nu>)^4>\over <\nu>^4}-3{(<\nu^2>-<\nu>^2)^2\over <\nu>^4}
\Big]-
18{<\nu^2>-<\nu>^2\over <\nu>^2}{\kappa_2\over<\nu>}+ $ \\
4&\\
  & $+3{<\nu^2>-<\nu>^2\over
<\nu>^2}\Big({\kappa_2\over<\nu>}\Big)^2+4{<\nu^2>-<\nu>^2\over
<\nu>^2}{\kappa_3\over<\nu>^2}+{\kappa_4\over<\nu>^3}$ \\
&\\
\hline
\end{tabular}
\end{Large}
\end{center}
\caption{The factorial cumulants $K_q$ for $q=2, 3$ and $4$, obtained
from
(~\protect\ref{9}) and (~\protect\ref{8}). When the fluctuations in the number
of strings are 
negligeable the cumulants are the cumulants resulting from $<\nu>$ independent
sources: $K_q=\kappa_q/<\nu>^{q-1}$. When string fluctuations dominate, as in
heavy nuclei collisions, only the terms inside square brackets are important.
In that limit, the factorial cumulants $K_q$ are given by the cumulants of the
string distribution. In that same limit the factorial moments $F_q$ are given
by the moments $<\nu^q>/<\nu>^q$ of the string distribution. For definition of
moments see, for instance, ~\protect\cite{16}.}

\end{table}

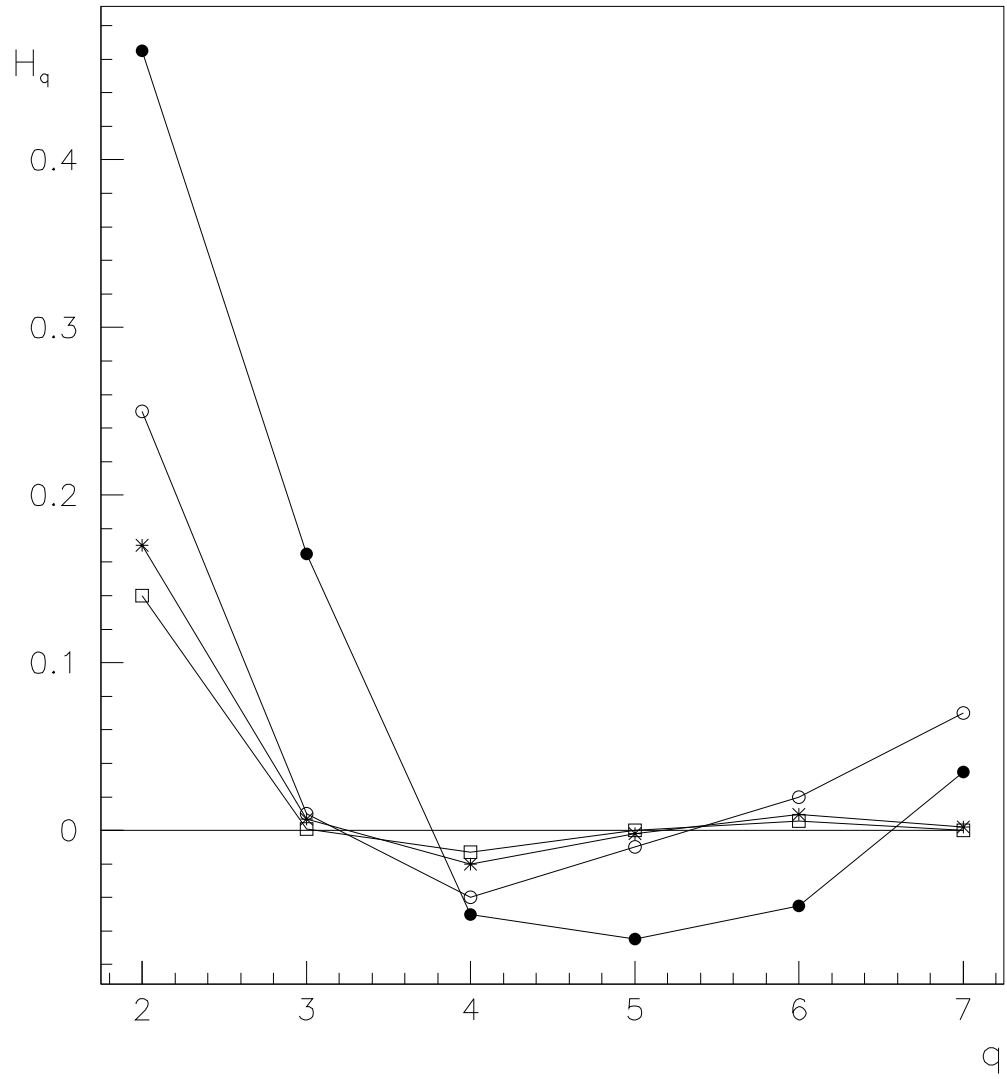
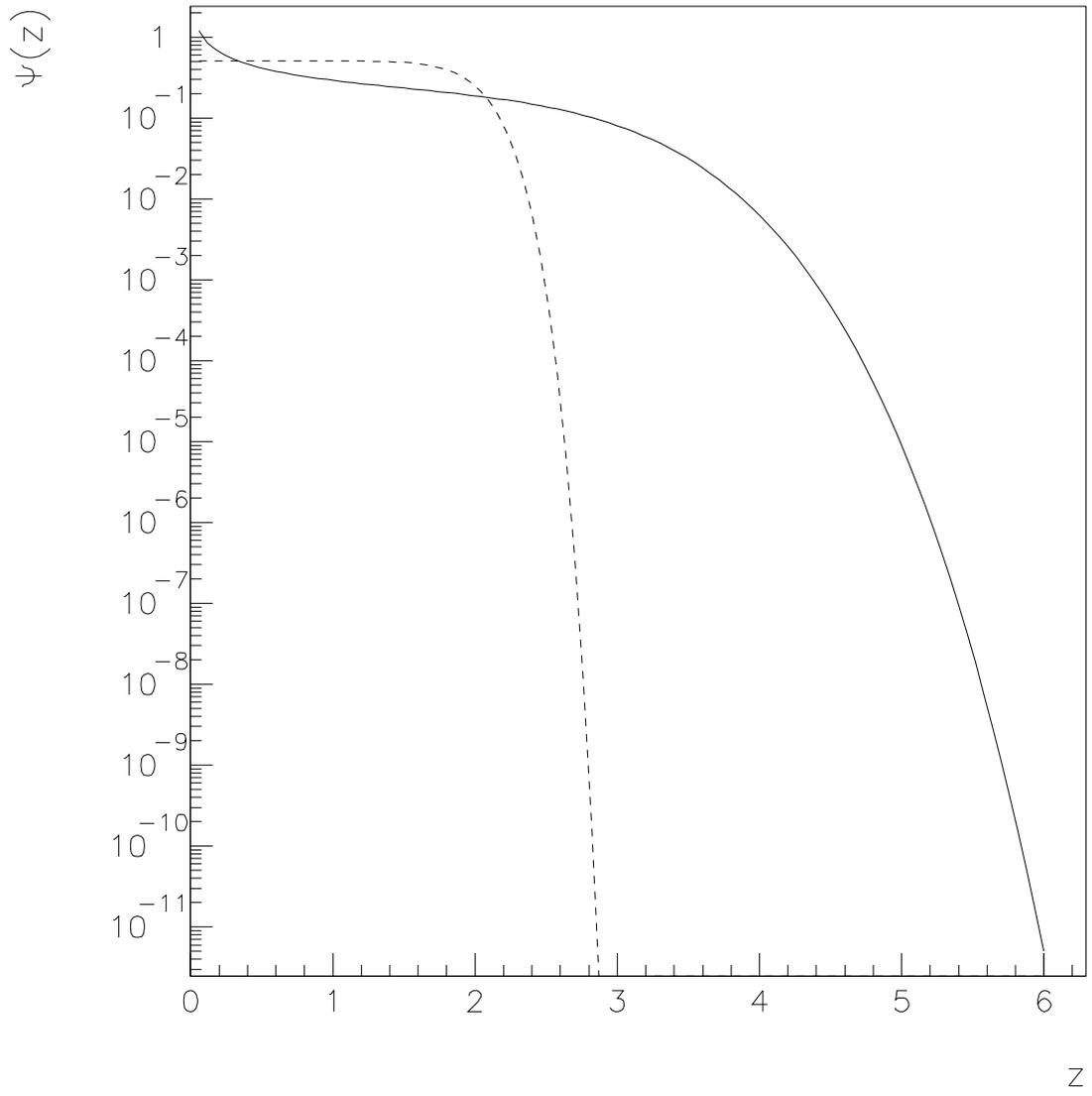
\end{document}